\documentclass[nofootinbib,twocolumn,aps,pre,superscriptaddress,citeautoscript,floatfix]{revtex4-2}
\usepackage[title]{appendix}
\usepackage{graphics}
\usepackage{graphicx}
\usepackage{color}
\usepackage{amsmath}
\usepackage{amssymb}
\usepackage{float}
\newcommand\be{\begin{equation}}
\newcommand\ee{\end{equation}}
\newcommand\bea{\begin{eqnarray}}
\newcommand\eea{\end{eqnarray} }
\newcommand{\kB }{k_{\rm B}}
\newcommand{\lB}{l_{\rm B}}
\newcommand{\lD}{\lambda_{\rm D}}
\begin{document}

\title{Charge Regulation of Polyelectrolyte Gels:\\ Swelling Transition}

\author{Bin Zheng}
\email{zhengbin@ucas.ac.cn}
\affiliation{Wenzhou Institute, University of Chinese Academy of Sciences, Wenzhou, Zhejiang 325000, China}
\affiliation{School of Physics and Astronomy,
	Tel Aviv University, Ramat Aviv 69978, Tel Aviv, Israel}
\author{Yael Avni}
\affiliation{James Franck Institute, University of Chicago, Chicago, IL 60637, USA}
\affiliation{Raymond and Beverly Sackler School of Physics and Astronomy,
Tel Aviv University, Ramat Aviv 69978, Tel Aviv, Israel}
\author{David Andelman}
\email{andelman@post.tau.ac.il}
\affiliation{Raymond and Beverly Sackler School of Physics and Astronomy,
Tel Aviv University, Ramat Aviv 69978, Tel Aviv, Israel}
\author{Rudolf Podgornik}
\email{podgornikrudolf@ucas.ac.cn. Also affiliated with the Department of Physics, 
Faculty of Mathematics and Physics, University of Ljubljana, 1000 Ljubljana, Slovenia}
\affiliation{School of Physical Sciences and Kavli Institute for Theoretical Sciences, 
University of Chinese Academy of Sciences, Beijing 100049, China}
\affiliation{Wenzhou Institute, University of Chinese Academy of Sciences, Wenzhou, Zhejiang 325000, China}
\affiliation{CAS Key Laboratory of Soft Matter Physics, Institute of Physics, Chinese Academy of Sciences, Beijing 100190, China}

\begin{abstract}
We study the effects of charge-regulated acid/base equilibrium on the 
swelling of polyelectrolyte gels, by considering a combination 
of the Poisson-Boltzmann theory and a two-site charge-regulation model based on the Langmuir adsorption isotherm. By exploring the volume change as a function of salt concentration for both nano-gels and micro-gels, we identify conditions  
where the gel volume exhibits a discontinuous swelling transition.
This transition is driven exclusively by the charge-regulation mechanism and is characterized by a 
closed-loop phase diagram. Our predictions can be tested experimentally for polypeptide gels.
\end{abstract}
\maketitle

\section{Introduction}
Polyelectrolyte (PE) hydrogels~\cite{Ferenc2021} are composed of chemically or physically cross-linked PE chains~\cite{Katch1955}, 
and are ubiquitous in 
soft bio-matter, being a quintessential part of the machinery of life \cite{Tanaka2004}. 
Apart from applications in controlled encapsulation for
drug release, hygiene products, sequestration of ions, and water desalination, 
they are used as raw materials 
for biomimetic actuators in soft robotics and therapeutic replacements~\cite{Thakur2018,Wilcox2021}. 
An important facet of PE hydrogel behavior is 
the large volume change exhibited by finite-size samples \cite{Khokhlov2002,Tokita2022,Man2021,Ding2022}. Such an effect is 
caused by a combination of electrostatic repulsion of charged monomers and osmotic 
pressure of their counterions~\cite{JiaDi2021,Muthu2017}. 
Most PE hydrogels contain weakly dissociable acidic or  
basic groups, which can change their ionization degree upon a change 
in the solution pH or salt concentration (see Ref.~\cite{Ferenc2021} for a recent review). The gel 
either releases or withholds 
protons to/from the bathing solution~\cite{Hofzum2021}, 
and its swelling is sensitive to small variations in these solution parameters.

The swelling of PE hydrogels can be either continuous or discontinuous, 
as first observed in experiments,  
and later rationalized by Tanaka and co-workers~\cite{Tanaka1984,Tokita2022} using the Donnan equilibrium theory. 
Those findings indicate that PE hydrogel swelling increases with pH, 
especially for pH values higher than the pK of ionizable polyacid groups. 
Furthermore, the swelling can become non-monotonic for high salt concentrations, 
first increasing at low salt, and then decreasing at higher salt concentrations. 

The theoretical analysis of the PE gel swelling was originally centered on the concept of Donnan equilibrium and gel elasticity~\cite{Tanaka1984,Tokita2022}. 
Later, it was extended~\cite{Ballauf2012a, Ferenc2021} to include a more detailed description of 
electrostatic interactions based either on the linear Debye-H\"uckel theory~\cite{English1998,Prausnitz2006},
or on the full non-linear Poisson-Boltzmann (PB) 
theory, combined with Flory-Huggins  
theory for polymer mixtures and gel elasticity~\cite{Levin2014,Nikam2020,Monica2011}. 

In previous models,  the gel charges were usually assumed to remain fixed. 
This constraint was lifted by Muthukumar {\it et al.}~\cite{Muthu2010,Muthu2012},  who introduced 
a {\it self-consistent charge regularization} of the polymer strands.
Later on, {\it charge regulation} (CR) was explicitly introduced into the PE gel theory by Szleifer {\it et al.}~\cite{Monica2012,Monica2014}, 
who described the ability of the PE chains to self-regulate their effective charge through coupling 
between ionization equilibrium and polymer conformations. A similar line of reasoning 
was used for PE chains with weakly dissociable groups~\cite{Drozdov2016, Kosovan2017, Jonas2019}. 
While the CR modeling in these works is limited to the Langmuir isotherm for a single type of dissociable sites, the 
general CR phenomenon~\cite{Rudi2018} encompasses also multiple and different types of dissociation sites.~\cite{borkovec2001ionization}.

In the present work, we consider both nano-gels and micro-gels, whose size is on the order of $100$nm and $1{\rm\mu m}$, 
respectively ~\cite{Ballauff2007,Potemkin2019}, and focus on the case of two types of dissociation sites on the PE chains. Specifically, the first type can become negatively charged by deprotonation, while the second can become positively charged by protonation~\cite{Monica2011a,Monica2012}. We present a mean-field reformulation based on the PB theory coupled to the free energy of a two-site CR model that 
connects self-consistently the ionic profiles and the PE dissociation state~\cite{Yael2019,Rudi2018,Huang2021}. 

We investigate the effect of salt at neutral pH on the swelling state of a spherical gel sample, by demanding the vanishing of the overall osmotic pressure after the addition of the elastic free energy, just as is done in the standard Flory-Rehner approach ~\cite{Levin2014}. We find an unexpected discontinuous 
volume transition as a function of added salt. 
Moreover, unlike the well-known swelling transition induced by the Flory-Huggins interaction parameter and analyzed {\it in extenso} 
in previous works~\cite{English1998}, the new discontinuous volume transition is due exclusively to the CR process itself. It stems 
directly from the 
acid-base equilibrium on the polyelectrolyte chains, and is somewhat akin to the critical behavior of CR macroions~\cite{Yael2020}. 
The corresponding gel charging {\it via} electrostatic coupling leads to a change in the swelling state, resulting in a phase diagram exhibiting either a 
continuous or a discontinuous swelling transition 
with upper and lower critical points. 

The outline of the paper is as follows. In section II, we describe the model and the corresponding free energy. Our results are then presented in section III,  where we show the electrostatic potential and polymer charge-density profiles, and analyze the gel swelling phase transition as a function of salt concentration and CR parameters. In section IV, we discuss our findings and make a comparison between the full PB theory and the Donnan approximation. 
Finally, we draw some conclusions relevant to future experiments.

\section{Model and free energy}

\begin{figure*}[t!]
\centering
{\includegraphics[width=0.4\textwidth,draft=false]{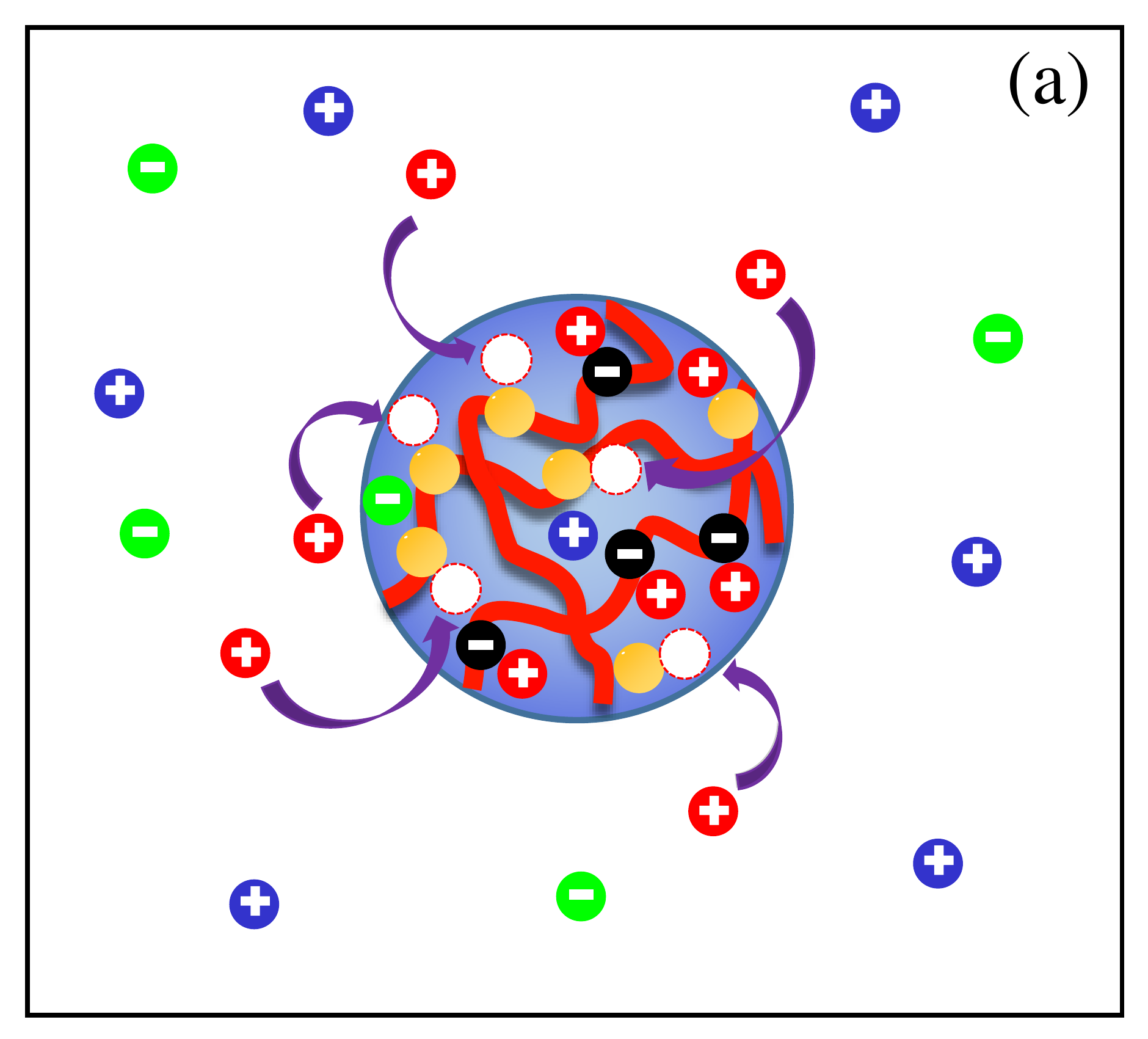}}
{\includegraphics[width=0.4\textwidth,draft=false]{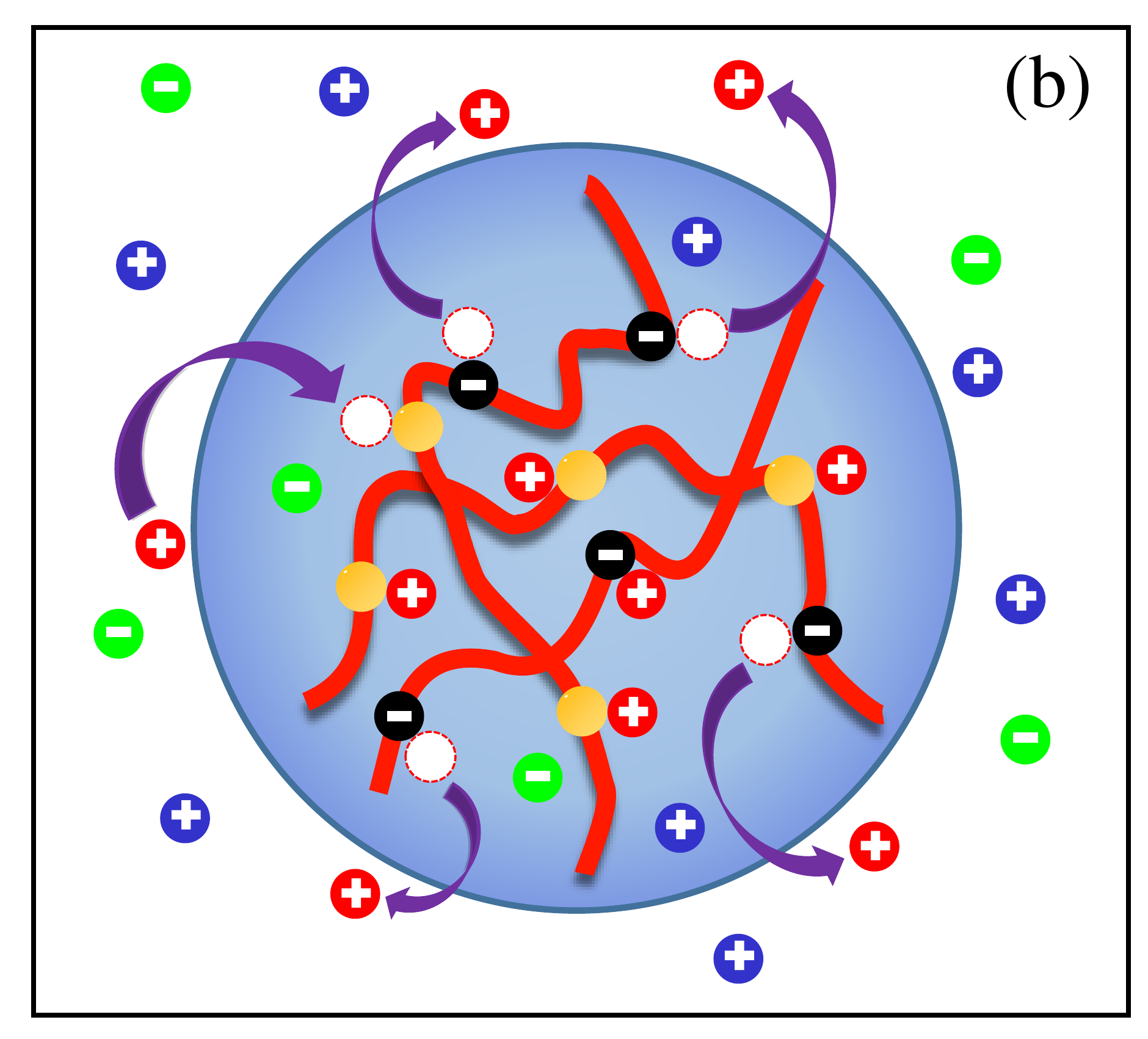}}
\caption{\textsf{Schematic presentation of two gel states. (a) collapsed state and (b) swollen states for a PE gel with protonation/deprotonation sites, formed by cross-linking $N_c$ polyelectrolyte chains. The gel volume is represented by the blue sphere. In the {\it two-site model},  the polymer chain has an equal number of $n$-type (deprotonation) and $p$-type (protonation) dissociation sites that can become negatively and positively charged, respectively. The ${\rm A}^{-}$ sites and ${\rm B}$ sites are represented by the black and yellow filled circles, respectively. All mobile ionic species, ${\rm H^+}$ (red-filled circles), anion (green-filled circles), and cation (blue-filled circles), are exchangeable between the gel volume and the bulk bathing solution.}}
\label{fig1}
\end{figure*}

We consider a gel composed of 
$N_c$ cross-linked chains of $N$ monomers each. The gel is immersed in a bathing $1$:$1$ 
solution containing monovalent ions of bulk concentration $n_{+}^b=n_{-}^b=n_b$, and ${\rm H^+}$ and ${\rm OH^-}$ ions with bulk concentrations $n^b_{\rm H}$ and $n^b_{\rm OH}$, respectively.
The gel chains contain two types of sites, a $p$-site undergoing a protonation,
and an $n$-site undergoing a deprotonation dissociation reaction.
The dissociable group in the $n$-type site is ${\rm AH}$, and is described by the reaction, 
\be
\begin{aligned}
\rm AH~~ \rightleftarrows ~~A^{-}+H^{+},
\end{aligned}
\ee
with the A-site becoming negatively charged in the process, after releasing a proton into the bath.
Similarly, the protonation of the $p$-type sites is characterized by an association of 
a proton from the bath
\be
\begin{aligned}
\rm B+H^{+}~~ \rightleftarrows ~~BH^{+} \ ,
\end{aligned}
\ee
implying that the B-site becomes positively charged. We assume that there are $N_{\rm site} = \gamma N $ 
chargeable sites in the gel, characterized by the fraction $\gamma$, $0 \leq \gamma \leq {\textstyle\frac12}$, and the sites are divided equally into $p$-type and $n$-type.

The gel is taken as a sphere of radius $R$, see Fig.~\ref{fig1}. Because of the radial symmetry, its properties depend only on the distance
 from the center, $r<R$.
While the concentration of sites is uniform inside the gel, we allow the degree of association/dissociation to vary locally.
Hence, we introduce the local charged site fractions, $\phi_{p}(r)$ and $\phi_{n}(r)$  for the p-type and n-type sites, respectively, 
that depend on the radial variable $r$.
$\phi_{n,p}$ varies from zero when the sites at radius $r$ are neutral to $1$ when the sites at radius $r$ are fully charged.
The fractions $\phi_{p}(r)$ and $\phi_{n}(r)$ are considered as thermodynamically annealed (adjustable) variables
that are coupled to the electrostatic potential profile $\psi(r)$.
Their thermodynamic equilibrium values are obtained by a  variational principle as is explained below.

\subsection{Equilibrium thermodynamics}
The thermodynamic potential $(G)$ contains several terms that can be written as a volume integral 
of the electrostatic free-energy density ($f_{\rm el}$), mobile ions translational entropy
density ($f_{\rm ent}$), and the CR free-energy density ($f_{\rm cr}$). 
Other terms stem from the chemical potentials and  
the gel elastic free-energy ($F_{\rm gel}$), 
\be
\begin{aligned}
	G&= \int_V ~g~ d^3 r +~F_{\rm gel} \\
   &= \int_Vd^3 r ~\Big(f_{\rm el}+f_{\rm ent}+f_{\rm cr}  \\
   &  - ~ \mu_-n_{-} - \mu_+n_{+} - \mu_{\rm H} n_{\rm H} - \mu_{\rm OH} n_{\rm OH}\Big) ~+~F_{\rm gel} \ ,
\label{energy1}
\end{aligned}
\ee
with $\mu_{\pm}$, $\mu_{\rm H}$ and $\mu_{\rm OH}$ being the chemical potentials of the cations, anions, ${\rm H^+}$ and ${\rm OH^-}$, respectively. 
The electrostatic free-energy density is given by~\cite{Safinya}
\be
\begin{aligned}
	&f_{\rm el} =-\frac{\varepsilon_0\varepsilon_r}{2}(\nabla\psi)^2 \\
	& + e\psi\left[n_+-n_-+n_{\rm H}-n_{\rm OH}- (\phi_n-\phi_p) \gamma c_{\rm gel} ~{\mathcal H}(R-r)\right] \ ,
\end{aligned}
\ee
where $n_{\pm}(r)$, $n_{\rm H}(r)$ and $n_{\rm OH}(r)$ are the local number densities of cations, anions, 
${\rm H^+}$ and ${\rm OH^-}$, respectively. 
The last term in the above equation corresponds to the net charge density of the polymer chains, 
where $c_{\rm gel}=N/V_{\rm gel}$ is the average gel concentration,
 $V_{\rm gel}=4\pi R^3/3$ is its volume, and ${\mathcal H}(x)$ is the Heaviside step function defined as ${\mathcal H}(x)=0$ for $x<0$ and
 ${\mathcal H}(x)=1$ for $x\ge 0$. 
 
The translational 
entropy on the mean-field level comes from all mobile ionic species and can be written in the ideal-gas form~\cite{Safinya}
\be
\begin{aligned}
& f_{\rm ent}= \kB  T \!\!\! \!\!\! \sum_{i=\pm, {\rm H, OH}} \!\!\! \!\!\! n_i\Big( \ln(n_ia^3)-1\Big) \ ,
\end{aligned}
\ee
where all ionic species are assumed to have the same molecular volume $a^3$. 
The CR term has been discussed extensively 
in Ref.~\cite{Yael2018} 
and can be written in a general form as  
\be
\begin{aligned}
f_{\rm cr}= \left[f_n\left( \phi_{n}\right)+f_p\left( \phi_{p}\right)\right]{\mathcal H}(R-r)\ ,
\end{aligned}
\ee
where a Langmuir adsorption is considered for the charge association/dissociation processes, corresponding to
\be
\begin{aligned}
f_p\left( \phi_{p} \right) &= (\alpha_{p} +\mu_{p}) z_p(\phi_{p})+ \gamma c_{\rm gel}s(\phi_{p} ), \\
f_n\left( \phi_{n} \right) &=(\alpha_{n} +\mu_{n}) z_n(\phi_{n})+ \gamma c_{\rm gel}s(\phi_{n} ),
\end{aligned}
\ee
where $\alpha_{{p, n}}$ are the association/dissociation free-energy changes for a single $p$- or $n$-site, respectively,  
with charge fractions $z_n(\phi_{n}){=}-\gamma c_{\rm gel}(1-\phi_{\rm n})$, and $z_p(\phi_{p}){=}-\gamma c_{\rm gel}\phi_{p}$, and 
$\mu_{{p, n}}=\mu_{\rm H}$ in this work, as we assume reactions are protonation/deprotonation processes. Finally, 
$s(\phi_{p, n})$ is the standard lattice-gas entropy
\begin{eqnarray}
s(x) = \kB T  [ x\ln x + (1-x) \ln(1-x)]. 
\end{eqnarray}

The gel elastic energy is computed based on the assumption of an affine deformation~\cite{Flory1953,Levin2014}  and depends on 
the actual model used for the polymer network~\cite{Landsgesell2021}. 
For an isotropic deformation, we assume $F_{\rm gel}$ to be of the form
\be
F_{\rm gel}=- {\textstyle\frac{3}{2}} \kB T N_{c} 
\Big(\ln{\frac{R}{R_{0}}}-\Big(\frac{R}{R_{0}}\Big)^2+1\Big),
\ee
where ${R}/{R_{0}}$ is the swelling ratio, with $R_{0}=\eta (N^{1/3} a)$ being a phenomenological 
{\it reference radius} parameter, proportional to the radius of the dry gel, in which the polymer sites 
are assumed to be in their close-packing configuration and $\eta$ is a phenomenological parameter.  

We now define several dimensionless variables, $ \tilde \psi \equiv {\it e}\psi/\kB T$, 
$\tilde\mu_i \equiv  \mu_i/\kB T$ for $i=\pm, {\rm H^+},{\rm OH^-}$ and $\tilde\alpha_{p, n} \equiv
\alpha_{p, n}/\kB T$ to be used hereafter, but for clarity purposes, we drop the tilde from all dimensionless variables. 

The thermodynamic equilibrium is then obtained by using a variational principle 
on the thermodynamic potential $G$ with respect to the annealed 
variables, $n_{\pm}$, $n_{\rm H}$, 
$\psi $ and $\phi_{p, n}$. In addition, we assume that the added salt concentration is much larger than ${\rm H}^+$, in which case $n_{\rm H}^b\ll n_b$. 
The variation with respect to $n_{\pm}$ and $n_{\rm H}$ yields
\begin{eqnarray}
n_{\pm}(\psi)\!=\!n_{b}{\rm e}^{\mp\psi}, ~ n_{\rm H}(\psi)\!=\!n^b_{\rm H}{\rm e}^{-\psi},
\end{eqnarray}
where $n_b=n^b_\pm=a^{-3}{\rm exp}(\mu_\pm)$ is the bulk salt concentration, taken at zero 
reference electrostatic potential $\psi=0$, and $n^b_{\rm H}=a^{-3}{\rm exp}(\mu_{\rm H})$. 

Variation with respect to $\psi$ yields the mean-field PB equation
\be
\nabla^2\psi=-4\pi \lB\Big(n_+(\psi)+n_{\rm H}(\psi)-n_-(\psi)- (\phi_{\rm n}-  
\phi_{p}) \gamma c_{\rm gel}{\mathcal H}(R-r)\Big) \ ,
\label{pb}
\ee
where $\lB={\it e}^2/(4\pi \varepsilon_0\varepsilon_r\kB T)$ is the Bjerrum length. 
For $n^b_{\rm H} \ll n_{b}$, the concentration of H$^+$
in the bulk is negligible as compared to the salt concentration, and the PB equation can be cast as
\be
\nabla^2\psi=\kappa^2\sinh{\psi}+4\pi \lB(\phi_{\rm n}-\phi_{p})\gamma c_{\rm gel}~{\mathcal H}(R-r) \ ,
\label{pb2}
\ee
where $\kappa^2 = 8\pi \lB\sqrt{(n_b+n^b_{\rm H})n_b }  \simeq 8\pi \lB n_b$, and $\kappa=1/\lD$
is the inverse Debye screening length. For details see footnote~\cite{Approx}. 

A further variation of the thermodynamic potential with respect to $\phi_{p, n}$ yields the dissociation equilibrium, 
$\phi_{p, n} = \phi_{p, n}(\psi)$, which looks like a Langmuir isotherm expression
\be
\begin{aligned}
& \phi_n=\frac{1}{(n^b_{\rm H} a^3) {\rm e}^{\alpha_n-\psi}+1} = \frac{1}{ {\rm e}^{\bar\alpha_n-\psi}+1} \ ,\\
& \phi_p=\frac{1}{(n^b_{\rm H} a^{3})^{-1} {\rm e}^{\psi-\alpha_p}+1} = \frac{1}{ {\rm e}^{\psi-\bar\alpha_p}+1} \ ,
\label{simplephirelation3}
\end{aligned}
\ee
where we defined the dimensionless parameters $\bar{\alpha}_{p,n}\equiv \alpha_{p,n}+\ln n^b_{\rm H} a^3$.

Finally, after introducing standard forms for the ${\rm pH}$ and the ${\rm pK}_{p, n}$, ${\rm pH} = - \log_{10}{(n^b_{\rm H} a^3)}$ and $\alpha_{p, n} = {\rm pK}_{p, n} \ln{10}$, 
we obtain a simple relation between the ${\rm pH}$ and the ${\rm pK}$'s with $\bar\alpha_{p, n}$ parameters
\be
\begin{aligned}
\bar\alpha_{p, n} \equiv -({\rm pH} - {\rm pK}_{p, n}) \ln{10}=\alpha_{p, n}-{\rm pH}\ln{10} \ .
\end{aligned}
\ee
Note that the above equation is cast in a form analogous to the {\it charge regulation} boundary condition in the 
Ninham-Parsegian model~\cite{NP-regulation}, except that  
the charge regulation is applied here inside the gel volume, and not on its surface.

The spatial electrostatic potential and charge profiles, $\psi = \psi(r)$ and $\phi_{p, n} = \phi_{p, n}(r)$, 
are obtained by solving numerically the PB equation, Eq.~(\ref{pb2}), 
and the dissociation equilibrium Eq.~(\ref{simplephirelation3}) in a spherical geometry, 
with the boundary conditions $\psi'(0)=0$ and $\psi'(\infty)=0$. Note that the electrostatic field at the center 
of the spherical gel and in the bathing solution has a zero value, dictated by symmetry and electro-neutrality in the bulk, respectively.

Finally, the total thermodynamic potential $G_{\rm tot}(R) =\int_V (g-g_{\rm b}) d^3 r+F_{\rm gel}$, where $g_{\rm b}={\rm const.}$ 
is the bulk thermodynamic potential density.
The equilibrium gel volume is obtained by 
minimizing $G_{\rm tot}(R)$ with respect to $R$, {\it i.e.}, by setting the osmotic pressure to zero.

\section{Results}

We consider two different gel sizes; a {\sl nano-gel} and a {\sl micro-gel}. The spherical nano-gel has
a radius $R = 100$\,nm and is made of 
$N_c=15$ cross-linked polymer chains. The spherical micro-gel, on the other hand, has a larger radius in the range of $0.5\leq R \leq 2\,{\rm \mu m}$ (same parameters as in the expression of Ref.~\cite{Levin2014}) and is made of $N_c=6\times10^3$ cross-linked chains.  
For both cases, each polymer chain consists of $N=500$  monomers, 
of which a fraction $\gamma=0.5$ contains dissociable groups. 
For simplicity, we assume that the monomers (sites on the chain) and ions 
have the same size,  $a=3.2~{\rm \AA}$. At room temperature and in water, the Bjerrum length is
 $\lB=7.2\,{\rm \AA}$, bulk {${\rm pH}=7$ is neutral (as in pure water),  and the 
bulk concentration of free ions is in the range, 0.01\,mM} $\le n_b\le$ 1\,mM. Finally, we take $\eta = 5$ in defining the phenomenological reference radius parameter for all the numerical examples.

Nano-gels and micro-gels are studied separately in order to explore the differences between the full 
PB description, based on an explicit solution of the PB equation and a more restrictive {\it Donnan} description, based on an assumption of a piece-wise constant electrostatic potential, being equal to the Donnan potential inside the gel and zero outside. The size difference between nano-gels and micro-gels
plays a crucial role in the comparison between the two approaches, which become indistinguishable for large enough ($R \gg \kappa^{-1}$) gel volumes (see Appendix A for details).

The Donnan potential is obtained by assuming a vanishing electrostatic potential outside the gel and enforcing the electroneutrality condition from the PB equation, Eq.~(\ref{pb2}), inside the gel with a constant potential $\psi_D$, that is then obtained as a solution of
\begin{eqnarray}
\kappa^2\sinh{\psi_D}+4\pi \lB\gamma c_{\rm gel} \Big(\frac{1}{ {\rm e}^{\bar\alpha_n-\psi_D}\!+\!1}-
\frac{1}{ {\rm e}^{\psi_D-\bar\alpha_p}\!+\!1}\Big)= 0 \ , ~~~~~~
\label{chneut}
\end{eqnarray}
stemming from the full PB equation, Eq.~(\ref{pb2}), with the charge regulation conditions, Eq.~(\ref{simplephirelation3}), included. However, the difference between the nano- and micro-gel is not restricted only to the quantitative differences between the electrostatic potential and charge density profiles near the gel-solvent interface. As we will show later on, the qualitative behavior of gel swelling is found to be always continuous for nano-gels but can be discontinuous for micro-gels, similar to the case of self-interacting gels with a Flory-Huggins interaction parameter ~\cite{Tanaka1984}. 

\subsection{Profiles of electrostatic potential and gel charge fraction}

\begin{figure*}[t!]
	\centering
	{\includegraphics[width=0.4\textwidth,draft=false]{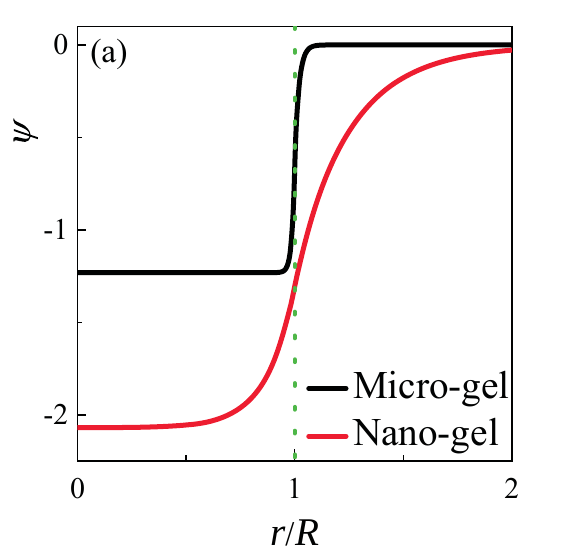}}
	{\includegraphics[width=0.4\textwidth,draft=false]{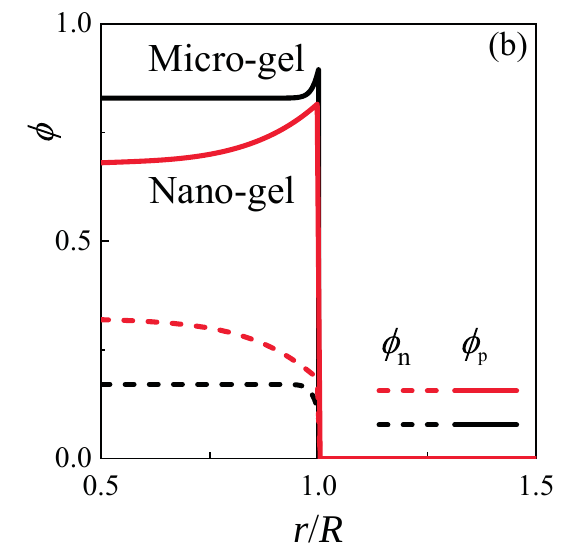}}
	\caption{
			\textsf{Profiles of (a) electrostatic potential $\psi(r)$, and (b) 
		gel dissociated charge fraction $\phi_{p, n}(r)$ as a function 
			of the dimensionless radial variable $r/R$ for nano- and micro-gels. The gel radius is $R=30\,{\rm nm}$ for the nano-gel and $R=0.5\,{\rm \mu m}$ for the micro-gel, $\bar\alpha_{p}=\bar\alpha_{n}=-2.3$ and the bulk salt concentration is $n_b=1{\rm mM}$, corresponding to Debye length of $\simeq 9.6\,{\rm nm}$. The dotted green line in (a) represents the location of the gel-solution interface.}
			}		
	\label{fig2}
\end{figure*}

In Fig.~\ref{fig2}a we present the electrostatic potential $\psi(r)$ and in 2b the gel dissociated charge fraction 
$\phi_{p, n}(r)$, as a function of $r$, the radial distance from the gel center.  
The nano-gel exhibits a strong variation inside the gel, while for the micro-gel, the electrostatic potential and dissociated charge fraction are almost constant inside and outside the gel, changing sharply only at the gel-solution interface, having a width comparable to the Debye screening length.
This suggests that the Donnan approximation implying a constant electrostatic potential throughout the gel region should work well for 
micro-gels but should be qualitatively incorrect for nano-gels. The relationship between the Donnan approximation and the full PB theory is further discussed in Appendix A. 

\subsection{The dependence of the swelling ratio and net charge on $\bar\alpha_{n}$}

\begin{figure*}[h]
\centering
{\includegraphics[width=0.4\textwidth,draft=false]{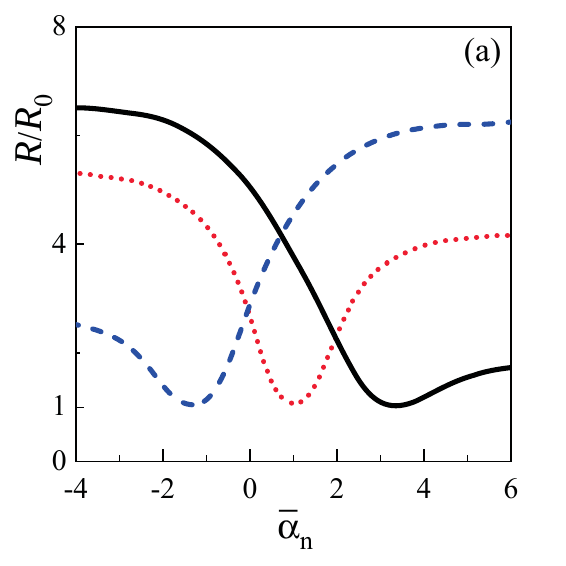}}
{\includegraphics[width=0.4\textwidth,draft=false]{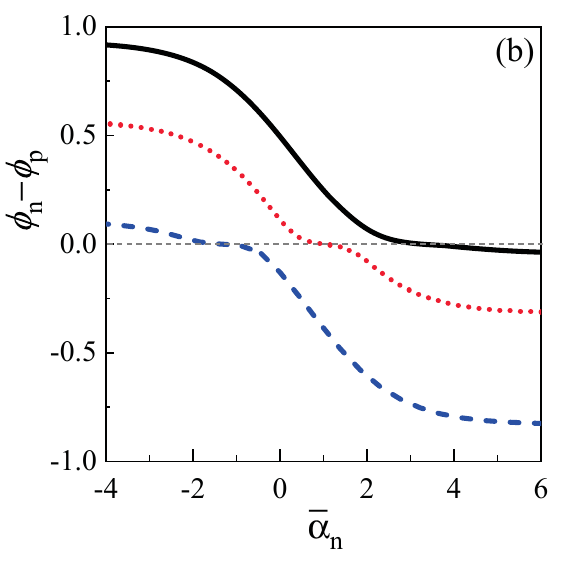}}
\caption{
\textsf{(a) Equilibrium swelling ratio $R/R_0$ for a nano-gel and (b) the corresponding total net charge, proportional to $\phi_n - \phi_p$, as a function of ${\bar\alpha_n}$, for ${\bar\alpha_p} = -2.3$ (black solid line), $0$ (red dotted line), and $2.3$ (blue dashed line) with $n_b=1\,{\rm mM}$. The position of the minimum in the swelling ratio varies as a function of ${\bar\alpha_p}$. The total net charge as a function of ${\bar\alpha_n}$ exhibits a polarity reversal going from positive to negative values through an  electroneutral state ($\phi_n=\phi_p$), coinciding with the position of the minimum in the swelling ratio in (a). 
The micro-gel case shows similar behavior (not shown here).}}
\label{fig3}
\end{figure*}

The dependence of the swelling ratio $R/R_0$ on $\bar\alpha_{n}=\alpha_{n}-{\rm pH}\ln{10}$, as shown in Fig.~\ref{fig3}, is non-monotonic and corresponds to different charging states of the gel. For fixed $\bar\alpha_{p}$, the swelling ratio displays a minimum, observable also in the opposite situation, where we fix $\bar\alpha_{n}$ and vary $ \bar\alpha_{p}$.

Depending on the values of $\bar\alpha_p$ and $\bar\alpha_n$, the gel can become overall positively or negatively charged as is seen in Fig.~\ref{fig3}. For large negative values of $\bar\alpha_{n}$, 
the gel is negatively charged with strong swelling. Increasing $\bar\alpha_{n}$ towards positive values causes first a gel contraction until it reaches a minimum size when the overall gel charge vanishes. A further increase in $\bar\alpha_{n}$ then further causes the gel to reverse its polarity and to become positively charged, 
as is seen in Fig.~\ref{fig3}b. Note that the magnitude of the swelling ratio for positive and negative polarity is different. 
This is a consequence of the charge asymmetry of the system because only ${\rm H}^+$ ions are exchanged with the gel, implying that negative and positive polarities are not equivalent. Similar behavior is also observed for the micro-gel case but is not shown here explicitly.

\subsection{The discontinuous swelling of micro-gels}

\begin{figure*}
{\includegraphics[width=0.4\textwidth,draft=false]{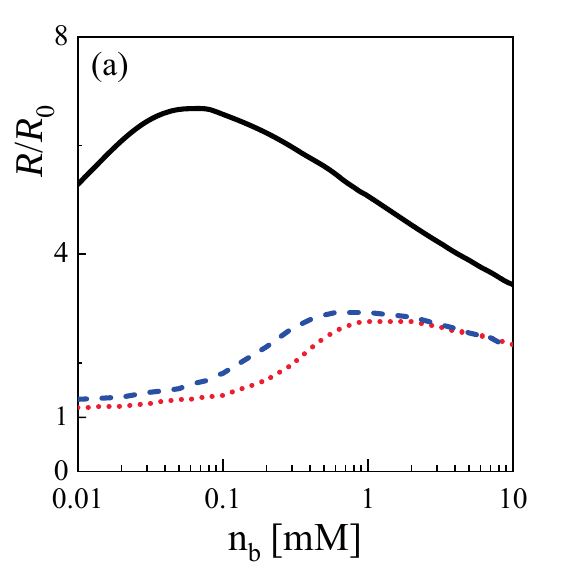}}
{\includegraphics[width=0.4\textwidth,draft=false]{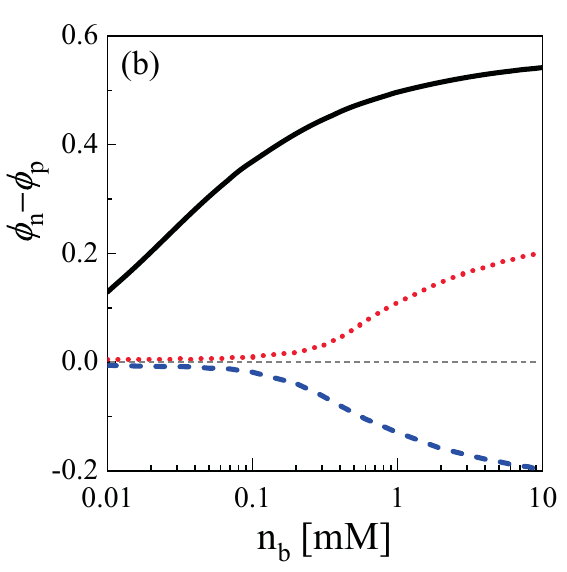}}
\caption{
\textsf{(a) Nano-gel equilibrium swelling ratio $R/R_0$, and (b) its corresponding net charge, proportional to  $\phi_n - \phi_p$,  as a function of the bulk salt concentration $n_b$ on a semi-log plot, for ${\bar\alpha_p} =-2.3$ (solid black line), $0$ (red dotted line), and $2.3$ (blue dashed line) for fixed $\bar\alpha_{n}=0$. The thin dashed line in (b) indicates the electroneutral state. The other parameters are $\gamma=0.5$, $\eta=5$ and $N=500$.}
}
\label{fig4}
\end{figure*}

\begin{figure*}
	{\includegraphics[width=0.4\textwidth,draft=false]{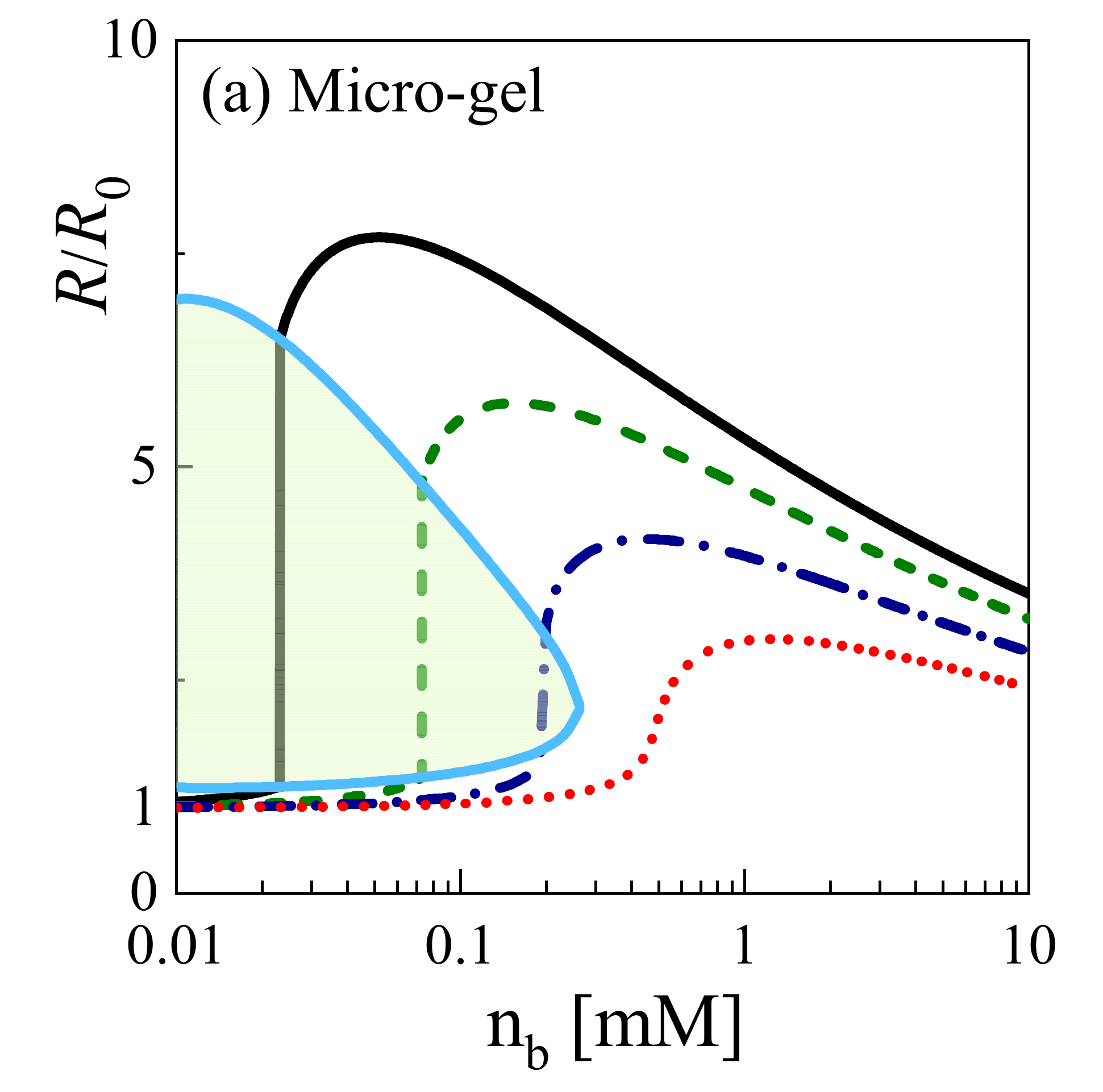}}
	{\includegraphics[width=0.4\textwidth,draft=false]{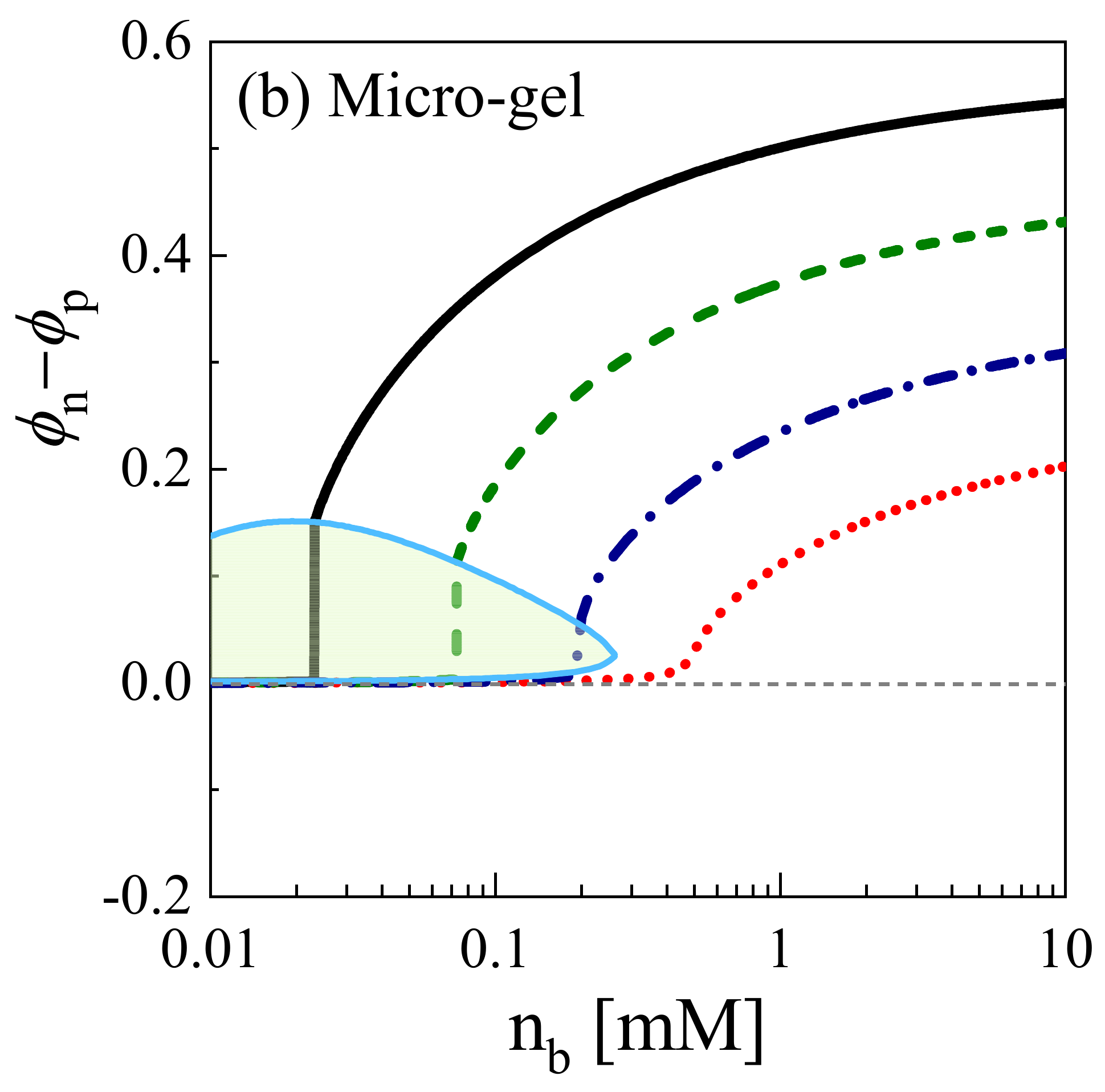}}
	\caption{
    	\textsf{(a) Equilibrium swelling ratio $R/R_0$ for a micro-gel and (b) its corresponding net charge, proportional to $\phi_n - \phi_p$, as a function of the salt concentration $n_b$ shown in a semi-log plot explicitly for several values of ${\bar\alpha_p}$: ${\bar\alpha_p} = -2.3$ (black solid line), $-1.15$ (green dashed line), $-0.46$ (dark blue dash-dotted line) and $0$ (red dotted line), for fixed $\bar\alpha_{n}=0$. A binodal line (light blue) separates in (a) the region that exhibits a discontinuous jump of $R/R_0$ (light green region) from the region that shows a continuous dependence on $n_b$. In (b) the region with a discontinuous jump in the net charge is separated from the region that shows a continuous dependence on $n_b$. The other parameters are $\gamma=0.5$, $\eta=5$ and $N=500$.}
	}
	\label{fig5}
\end{figure*}

Before discussing the interesting case of micro-gel, we show in figure~\ref{fig4}a the dependence of the swelling ratio $R/R_0$ on the bulk salt concentration $n_b$ 
for nano-gels. Generally speaking, $R/R_0$ displays a non-monotonic dependence on $n_b$, as the gel size first increases and then decreases. 
For small $n_b$ values, increasing $n_b$ will increase the dissociation of the {\rm AH} 
sites and, consequently, leading to an interesting observation of {\it charge-driven swelling} of the gel. 
As $n_b$ increases even further, the gel charge density gradually reaches a maximal value, 
while the electrostatic interactions get screened 
and the osmotic pressure difference between the external bathing solution and the gel decreases, causing a subsequent 
{\it charge-driven shrinkage} of the gel~\cite{Landsgesell2021}. Figure~\ref{fig4}b shows that the absolute value of the corresponding net charge of the nano-gel is an increasing function of $n_b$ and its polarity depends on the sign of ${\bar\alpha_p}$.

A very different scenario is seen for micro-gels for combinations of ($\bar\alpha_{p}, \bar\alpha_{n}$) values as shown in Fig.~\ref{fig5}a and \ref{fig5}b. Here, the $n_b$ dependence is not everywhere continuous but manifests a discontinuous change (a jump) of the swelling ratio as well as the net charge (see Appendix B for details).  In fact, the $R/R_0$ dependence on $n_b$ is quite similar to the {\sl van der Waals isotherms} ($P=P(V)$ for constant $T$) for the liquid-gas transition, where here the ``isotherms" correspond to certain combinations of the parameters $\bar\alpha_{p}$ and $\bar\alpha_{n}$. 

For small values of $n_b$ and when $\bar\alpha_n = 0$,  the positive and negative charge fractions are approximately equal irrespective of the $\bar\alpha_p$ value, resulting in an uncharged gel. As ${n}_b$ increases,  the micro-gel becomes charged 
with a polarity that depends on the sign of $\bar\alpha_p$. For $\bar\alpha_p \leq 0$ shown in Fig.~\ref{fig5}a and \ref{fig5}b, the negative charge fraction prevails and the polarity of the gel charge is overall negative. Upon increase of $n_b$, both the charge as well as the swelling ratio start continuously to increase, as seen in Fig.~\ref{fig5}a. This increase persists until at some large enough value of $n_b$ the micro-gel discontinuously charges up and expands by a several-fold increase in the swelling ratio. If $n_b$ is increased even further after the discontinuous jump the charge of the gel continues to grow but the swelling ratio first reaches a maximum and then starts decreasing just as in the nano-gel case. The existence of the discontinuous jump in the swelling ratio driven solely by charge dissociation and electrostatic interactions is one of our main results and sets the behavior of micro-gels quite apart from that of nano-gels.

The discontinuous jump in the charge and swelling ratio of the gel is not universal but depends on $\bar{\alpha}_p$ and $\bar{\alpha}_n$. For example, having $\bar{\alpha}_n=0$ and $\bar{\alpha}_p>0$ leads to continuous swelling and positive polarity (see Appendix C for details). A ``binodal line" in the ($R/R_0$, $n_b$) ``phase diagram" exhibits a critical point (Fig.~\ref{fig5}a) that separates the region of continuous from discontinuous charging and swelling behavior in the asymmetric charge-regulated PE gels. In fact, the region with discontinuous charging and swelling jumps displays a closed loop in the $(\bar{\alpha}_p, \bar{\alpha}_n)$ plane, as is clear from Fig.~\ref{fig6}. Invoking an analogy with a van der Waals gas one could associate the swelling ratio with pressure, $n_b$ with volume, and the $(\bar{\alpha}_p, \bar{\alpha}_n)$ combination with temperature.

\subsection{Closed-loop phase diagram of gel swelling}

\begin{figure}[t!]
	\centering
	{\includegraphics[width=0.4\textwidth,draft=false]{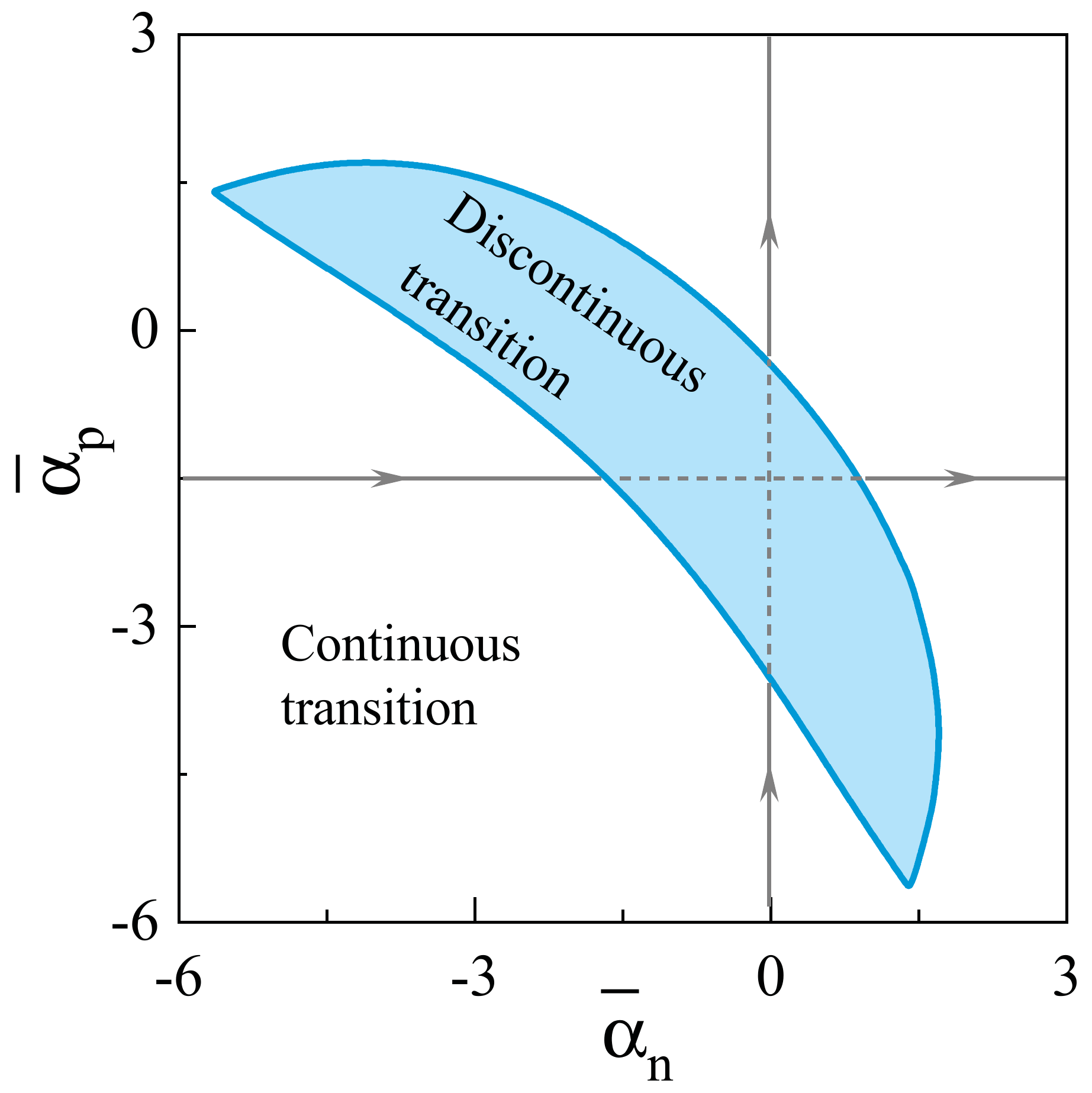}}
	\caption{\textsf{Closed-loop phase diagram of the micro-gel in the $({\bar \alpha_n}, {\bar\alpha_p})$ parameter space.
	The phase diagram is calculated from the $R/R_0$ dependence on $n_b$ via the Donnan approximation, and exhibits a closed-loop, delimiting the discontinuous transition (inner blue region) from the continuous one (outer region). The vertical and horizontal lines are examples of a re-entrant transition through the closed-loop region. On the vertical line ${\bar \alpha_n}$ is being kept constant and ${\bar\alpha_p}$ increases. The role of ${\bar\alpha_p}$ and ${\bar\alpha_n}$ is reserved on the horizontal line.}
}
	\label{fig6}
\end{figure} 

Referring to the continuous and the discontinuous gel swelling transitions, it would be important to understand the nature of the corresponding phase diagram in the $(\bar\alpha_{p}, \bar\alpha_{n})$ parameter space, as it gives a better global picture. 
In Fig.~\ref{fig6}, we show a {\it closed-loop} phase diagram where the swelling/charging transition is discontinuous inside the loop, 
while in the outside region, it is continuous. This means that at a fixed 
$\bar\alpha_{p}$ (or $\bar\alpha_{n}$), while increasing $\bar\alpha_{n}$ (or $\bar\alpha_{p}$), the gel first exhibits a continuous transition, then a discontinuous transition, and finally, 
it reverts to a continuous transition. Hence, the gel 
swelling transition is a {\it re-entrant phase transition}, in terms of the $\bar\alpha_{n}$ and $\bar\alpha_{p}$ system parameters.

It is important to stress 
that these effects are purely mean-field in nature and are due 
to the charge regulation and gel elasticity in charge asymmetric PE gels. 
They invoke a change in the gel polarity without having to invoke either a Flory-Huggins parameter, 
nor steric interactions and correlation effects as is the case in the Voorn-Overbeek type 
theories~\cite{Muthu2010,Muthu2012,Larson2016,Drozdov2016,Jonas2019,Landsgesell2021,Zheng2021}.

\section{Discussion and conclusions}

Charge regulation effects in swelling phase-transitions and charge separation phenomena of polyelectrolyte (PE) 
gels and solutions have been considered in the past. Additional related effects such as gel elasticity, the entropy of mixing, counterion adsorption, local dielectric constant, electrostatic interaction among polymer segments, and salt ion correlations have been also explored by analytical theories ~\cite{Muthu2010,Muthu2012} and extensive simulations ~\cite{Jonas2019,Landsgesell2021}. 

However, our main finding is the existence of a continuous/discontinuous swelling phase-transition of micro-gels, driven exclusively by the charge state equilibrium (Fig.~\ref{fig5}). Such swelling phase transition is not a consequence of the Flory-Huggins interaction parameter but is connected purely 
with the changes in the charging equilibrium as described by the CR mechanism. We also find that not every charge 
regulation model results in a re-entrant transition as is seen in the closed-loop phase diagram of Fig.~\ref{fig6}. 
The necessary ingredient of the model is the presence of both positively and negatively chargeable subunits, as well as a sufficiently large gel size that could be characterized as a micro-gel. 

The non-monotonic variation of the swelling 
ratio as a function of the salt concentration in the bathing solution then follows from the gel charge reversal is related to the charge 
regulation process. The finite size of nano-gels smooths the salt dependence of the swelling ratio, 
while for a large enough volume (micro-gels), its swelling becomes discontinuous and results in two macroscopic phases. 

The relation between charge regulation, as quantified by the two adsorption parameters $({\bar \alpha_n}$ and ${\bar\alpha_p})$, and electrolyte screening, as quantified by the 
salt concentration ($n_b$), then engender a closed-loop phase diagram with an inner region corresponding to a discontinuous swelling. We 
do not specifically investigate the coupling between the pure CR mechanism and possible non-electrostatic interactions 
as quantified by the Flory-Huggins interaction parameter that would result in a more complicated parameter space. 
The results presented above imply that CR effects pertinent to a PE gel with positively and negatively chargeable 
subunits can lead to an interesting scenario of swelling phenomena, quite different from what is known for the usual polycationic/polyanionic gels.

With the relation between $\bar\alpha_p, \bar\alpha_n$, and the pH and pK$_a$ of the protonation/deprotonation sites, one can reconstruct the swelling phase-diagram using those experimental and more accessible variables. This allows a better understanding of possible experimental realizations of our model, for example, a polypeptide gel, composed of cross-linked (oligo)peptide chains. Peptide amino acids are one of the most important examples of charge regulation, and their acid/base equilibria have been demonstrated in a variety of other phenomena. We hope that our model will motivate such experimental investigations in the future.

\vskip 2truecm
\noindent{\bf Acknowledgments}

\noindent
RP acknowledges funding from the Key Project No. 12034019 of the 
National Natural Science Foundation of China (NSFC) and the hospitality of 
the School of Physics, University of Chinese Academy of Sciences (UCAS) and 
Tel Aviv University where this work was completed.
DA acknowledges the NSFC-ISF Research Program, jointly funded by the 
NSFC under grant No.~21961142020 and the Israel Science Foundation (ISF) 
under grant No.~3396/19, and ISF grant No.~213/19. 
BZ acknowledges the NSFC through 
Grants (No.~22203022), and the Scientific Research Starting Foundation of Wenzhou Institute,  
UCAS (No.~WIUCASQD2022016). 
YA is thankful for the support of the Clore Scholars Programme of the Clore Israel Foundation and the Zuckerman STEM Leadership Program.
\newpage

\appendix

\section{Comparison of PB and Donnan approaches}

\begin{figure}[H]
	\centering
	{\includegraphics[width=0.4\textwidth,draft=false]{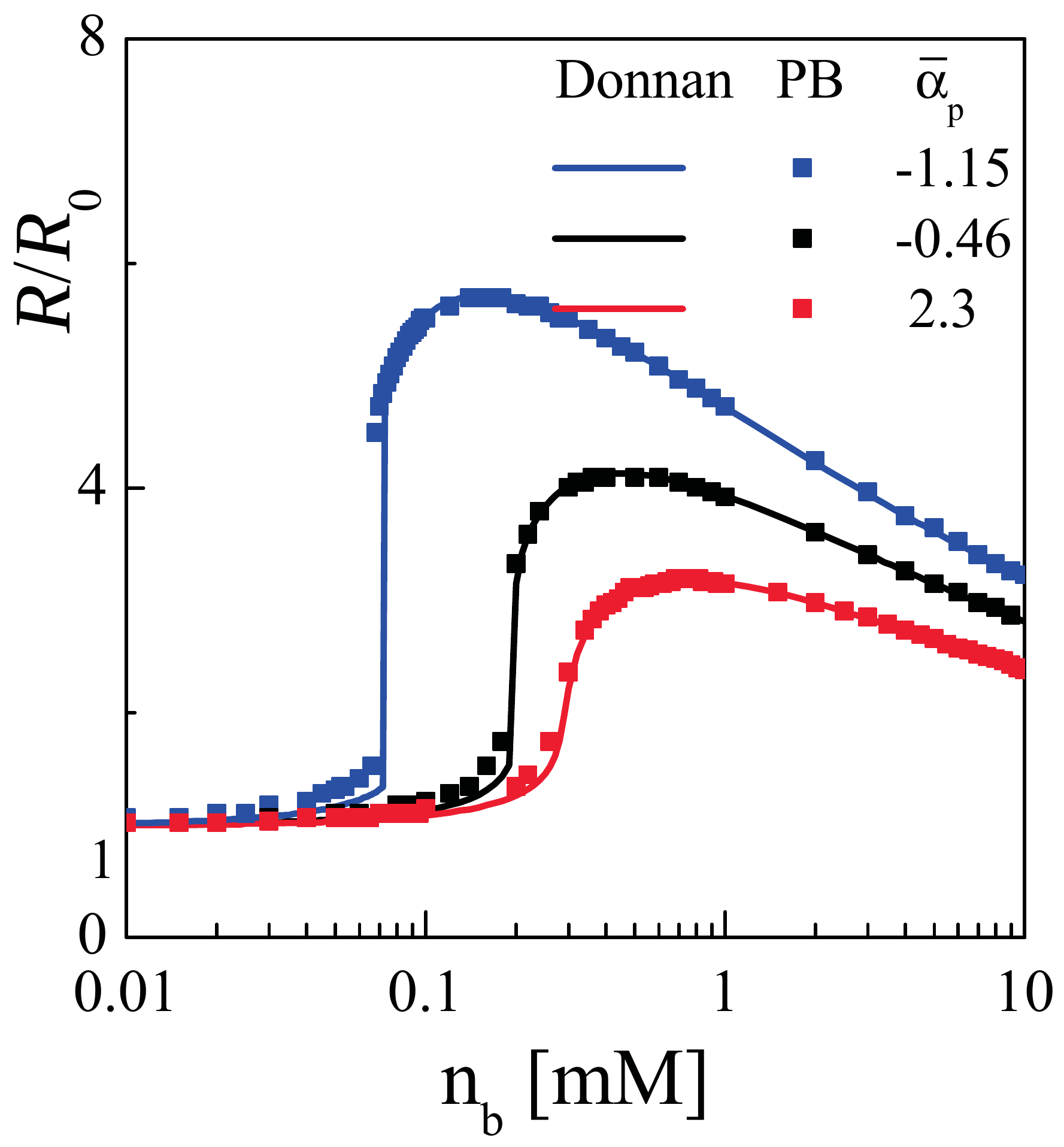}}
	\caption{
		\textsf{Comparison of the Donnan theory (solid line) and full PB theory (square symbols) for a micro-gel. The gel swelling transition is shown on a semi-log plot as a function of $n_b$ for different values of $\bar\alpha_{p}=-1.15, -0.46$ and $2.3$ while fixing $\bar\alpha_{n} = 0$.  The dependence of the swelling ratio $R/R_0$ on the bulk salt concentration $n_b$ given by the Donnan approximation is almost the same as the one given by the full PB theory for micro-gels.}}
	\label{figA1}
\end{figure}

For micro-gels, we compare in Fig.~\ref{figA1} the equilibrated swelling ratio $R/R_0$ calculated from the full PB theory with the Donnan approximation that assumes a piece-wise constant $\psi$ (namely, having a discontinuity at the gel interface). Inside the gel, $\psi$ is a constant corresponding to the Donnan potential, having a value that can be derived from the charge neutrality relation, Eq. (\ref{chneut}), that follows from the PB equation.

For $\bar\alpha_n=0$ and $\bar\alpha_p \leq 0$, the swelling of the gel expressed by $R/R_0$ is discontinuous, corresponding to a first-order swelling transition. The first-order phase transition is observed in the Donnan approximation as well as in the full PB theory, as shown in Figs.~\ref{figA1} and \ref{fig5}a. The numerical results of the two approaches are almost the same.

\section{Discontinuous gel swelling transition}

\begin{figure}[H]
	\centering
	{\includegraphics[width=0.4\textwidth,draft=false]{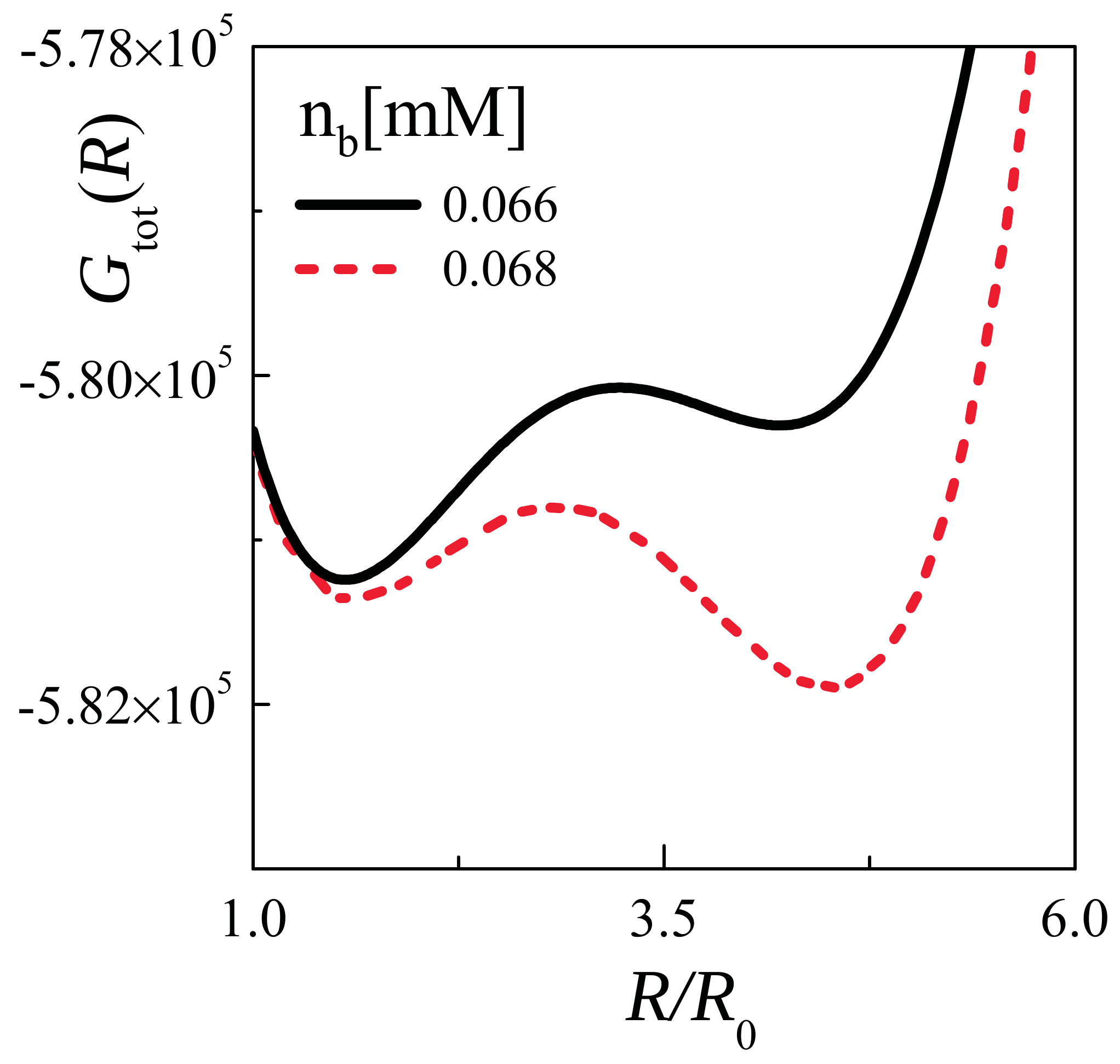}}
	\caption{
		\textsf{The total thermodynamic potential $G_{\rm tot}$ of a micro-gel as a function of the swelling ratio $R/R_{0}$, 
		for two values of salt concentration, $n_b=0.066$\,mM and $0.068$\,mM. Other parameters are $\bar\alpha_p=-1.15$ and $\bar\alpha_n=0$. 
	}}
	\label{figA2}
\end{figure}

The first-order phase transition of the two-site charge-regulation model within this work can be demonstrated by plotting the dependence of the thermodynamic potential on the swelling ratio $R/R_0$ for $\bar\alpha_n=0$ and $\bar\alpha_p \leq 0$. This is seen in Fig.~\ref{figA2} where two local minima correspond to the collapsed and swollen states. 
When the salt concentration $n_b$ changes from $0.066$\,mM to $0.068$\,mM, 
the global minimum of the thermodynamic potential jumps from the collapsed to the swollen state. This jump is an example of the jumps of the swelling gel ratio as seen in Fig.~\ref{fig5}a. 

\section{Continuous gel swelling transition}

\begin{figure*}
 {\includegraphics[width=.45\textwidth,draft=false]{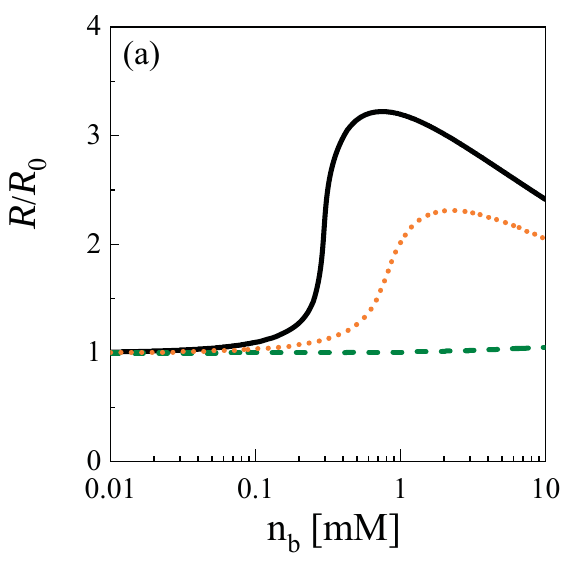}}
 {\includegraphics[width=.45\textwidth,draft=false]{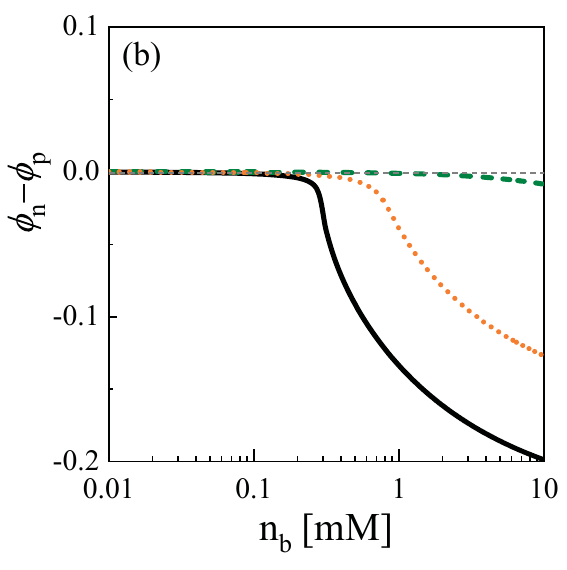}}
 \caption{
 \textsf{(a) The micro-gel equilibrium swelling ratio $R/R_0$ and (b) the corresponding net charge, proportional to  $\phi_n - \phi_p$, as a function of the salt concentration $n_b$ on a semi-log plot for $\bar\alpha_{n}=0$, with ${\bar\alpha_p} = 1.15$ (green dashed line), $1.84$ (orange dotted line), and $2.3$ (black solid line). For the choice of
 the $\bar\alpha_n$ and $\bar\alpha_p$ parameter value, the gel swelling is continuous but non-monotonic.
 } }
 \label{figA3}
\end{figure*}

For another choice of $\bar\alpha_n$ and $\bar\alpha_p$ such as $\bar\alpha_n=0$ and $\bar\alpha_p \geq 0$, the swelling of the micro-gel is continuous and the net charge of the gel shows no polarity change. An example of $\bar\alpha_{n}=0$ and $\bar\alpha_p = 1.15$ is shown in Fig. \ref{figA3}, where the gel swelling as a function of $n_b$ is continuous and shows an increase followed by a decrease, with a hump depending on the value of  $\bar\alpha_p$.

\newpage

\clearpage


\begin{thebibliography}{99}

\bibitem{Ferenc2021} Horkay, F. Polyelectrolyte gels: a unique class of soft materials. 
\textit{Gels} $\bf{2021}$, \textit{7}, 102.

\bibitem{Katch1955} Katchalsky, A.; Michaeli, I. Polyelectrolyte gels in salt solutions. 
\textit{J. Polym. Sci.} $\bf{1955}$, \textit{15}, 69-86.

\bibitem{Tanaka2004} Tanaka, T. \textit{From gels to life}; University of Tokyo Press, 2004.

\bibitem{Thakur2018} \textit{Polymer gels perspectives and applications}; Thakur, V. K.; Thakur, M. K.; Voicu, S. I., Eds.; Springer, 2018.

\bibitem{Wilcox2021}  Wilcox, K. G.; Kozawa, S. K.; Morozova, S. Fundamentals and mechanics of 
polyelectrolyte gels: thermodynamics, swelling, scattering, and elasticity. \textit{Chem. Phys. Rev.} $\bf{2021}$, {\textit 2}, 041309. 

\bibitem{Tokita2022} Tokita, M. Phase transition of gels - a review of Toyoich Tanaka's research, \textit{Gels} $\bf{2022}$, {\textit 8}, 550.

\bibitem{Man2021} Man, X. K.; Doi, M. swelling dynamics of a disk-shaped gel, \textit{Macromolecules} $\bf{2021}$, {\textit 54}, 4626–4632.

\bibitem{Ding2022} Ding, Z. Y.; Lyu, P. H.; Shi, A.; Man, X. K.; Doi, M. Diffusio-mechanical theory of gel bending induced by liquid penetration, \textit{Macromolecules} $\bf{2022}$, {\textit 55}, 550.


\bibitem{Khokhlov2002} Osada, Y; Khokhlov, A. \textit{Polymer gels and networks}, CRC Press, 2001.

\bibitem{JiaDi2021} Jia, D.; Muthukumar, M. Theory of charged gels: swelling, elasticity, and dynamics. \textit{Gels} $\bf{2021}$, {\textit 7}, 49.

\bibitem{Muthu2017} Muthukumar, M. 50th anniversary perspective: a perspective on polyelectrolyte solutions. 
\textit{Macromolecules} $\bf{2017}$, {\textit 50}, 9528-9560.

\bibitem{Hofzum2021} Hofzumahaus, C.; Strauch, C.; Schneider, S. Monte Carlo simulations of weak polyampholyte microgels: 
pH-dependence of conformation and ionization.  \textit{Soft Matter} $\bf{2021}$, {\textit 17}, 6029-6043.

\bibitem{Tanaka1984} Ricka, J.; Tanaka, T. Swelling of ionic gels: quantitative performance of the Donnan theory. 
\textit{Macromolecules} $\bf{1984}$, \textit{17}, 2916-2921.

\bibitem{Ballauf2012a} Yigit, C.; Welsch, N.; Ballauff, M.; Dzubiella, J. Protein sorption to charged microgels: characterizing 
binding isotherms and driving forces. \textit{Langmuir} $\bf{2012}$, \textit{28}, 14373-14385.

\bibitem{English1998} English, A. E.; Tanaka, T; Edelman, E. R. Polyampholytic hydrogel swelling transitions: limitations of 
the Debye-H\"uckel Law. \textit{Macromolecules} $\bf{1998}$, \textit{31}, 1989-1995.

\bibitem{Prausnitz2006} Victorov, A.; Radke, C.; Prausnitz, J. Molecular thermodynamics for swelling of a mesoscopic ionomer gel 
in 1:1 salt solutions. \textit{Phys. Chem. Chem. Phys.} $\bf{2006}$, \textit{8}, 264-278. 

\bibitem{Levin2014} Colla, T.; Likos, C. N.; Levin, Y. Equilibrium properties of charged microgels: a Poisson-Boltzmann-Flory approach. 
\textit{J. Chem. Phys.}  $\bf{2014}$, \textit{141}, 234902.

\bibitem{Nikam2020} Nikam, R.; Xu, X.; Kandu\v c, M.; Dzubiella, J. Competitive sorption of monovalent and divalent ions by highly charged 
globular macromolecules. \textit{J. Chem. Phys.} $\bf{2020}$, \textit{153}, 044904.

\bibitem{Monica2011} Jha, P. K.; Zwanikken, J. W.; Detcheverry, F. A.; de Pablo, J. J.; Olvera de la Cruz, M. Study of volume phase 
transitions in polymeric nanogels by theoretically informed coarse-grained simulations. \textit{Soft Matter} $\bf{2011}$, \textit{7}, 5965-5975.

\bibitem{Muthu2010} Muthukumar, M.; Hua, J.; Kundagrami, A. Charge regularization in phase separating polyelectrolyte solutions. 
\textit{J. Chem. Phys.} $\bf{2010}$, \textit{132}, 084901.

\bibitem{Muthu2012} Hua, J.; Mitra, M. K.; Muthukumar, M. Theory of volume transition in polyelectrolyte gels 
with charge regularization. \textit{J. Chem. Phys.} $\bf{2012}$, \textit{136}, 134901.


\bibitem{Monica2012} Longo, G. S.; Olvera de la Cruz, M.; Szleifer, I. Molecular theory of weak 
polyelectrolyte thin films. \textit{Soft Matter} $\bf{2012}$, \textit{8}, 1344-1354.

\bibitem{Monica2014} Longo, G. S.; Olvera de la Cruz, M.; Szleifer. I. Non-monotonic swelling of surface 
grafted hydrogels induced by pH and/or salt concentration. 
\textit{J. Chem. Phys.} $\bf{2014}$, \textit{141}, 124909.

\bibitem{Jonas2019} Landsgesell, J.; Sean, D.; Kreissl, P.; Szuttor, K.; Holm.C. Modeling gel swelling equilibrium in 
the mean-field: from explicit to Poisson-Boltzmann models. \textit{Phys. Rev. Lett} $\bf{2019}$, \textit{122}, 208002.

\bibitem{Drozdov2016} Drozdov, A. D.; Christiansen, D. J. The effects of pH and ionic strength of swelling of cationic gels. 
\textit{Int. J. Appl. Mech.} $\bf{2016}$, \textit{8}, 1650059.

\bibitem{Kosovan2017} Rud, O.; Richter, T.; Borisov, O.; Holm, C.; Ko\v sovan, P. A self-consistent mean-field model for polyelectrolyte gels. 
\textit{Soft Matter} $\bf{2017}$, \textit{13}, 3264-3274.

\bibitem{Rudi2018} Podgornik, R. General theory of charge regulation and surface differential capacitance. 
\textit{J. Chem. Phys.} $\bf{2018}$, \textit{149}, 104701.

\bibitem{borkovec2001ionization} Borkovec, M.; J\"{o}nsson, B.; Koper, G. J. M. Ionization processes and proton 
binding in polyprotic systems: small molecules, proteins, interfaces, and polyelectrolytes. In \textit{Surf. Colloid Sci.}, ed.; Springer Science, LLC, 2001; pp 99-339.

\bibitem{Ballauff2007} Ballauff, M.; Lu, Y. ``Smart" nanoparticles: Preparation, characterization, and applications. 
\textit{Polymer} $\bf{2007}$, \textit{48}, 1815-1823.

\bibitem{Potemkin2019} Karg, M.; Pich, A.; Hellweg, T.; Hoare, T.; Lyon, L. A.; Crassous, J. J.; Suzuki, D.; Gumerov,  R. A.; Schneider, S.; Potemkin, I. I.; Richter, W. Nanogels and microgels: from model colloids to applications, recent developments, 
and future trends. \textit{Langmuir} $\bf{2019}$, \textit{35}, 6231-6255.

\bibitem{Monica2011a} Longo, G. S.; Olvera de la Cruz, M.; Szleifer. I. Molecular theory of weak polyelectrolyte 
gels: the role of pH and salt concentration. \textit{Macromolecules} $\bf{2011}$, \textit{44}, 147-158.

\bibitem{Yael2019} Avni, Y.; Andelman, D.; Podgornik, R. Charge regulation with fixed and mobile charged macromolecules. 
\textit{Curr. Opin. Electrochem.} $\bf{2019}$, \textit{13}, 70-77.

\bibitem{Huang2021} Huang, C. H.; Zhu, Y. Y.; Man, X. K. Block copolymer thin films. 
\textit{Phys. Rep.} $\bf{2021}$, \textit{932}, 1-3.

\bibitem{Yael2020} Avni, Y.; Podgornik, R.; Andelman, D. Critical behavior of charge-regulated macro-ions. \textit{J. Chem. Phys.} $\bf{2020}$, \textit{153}, 024901.

\bibitem{Safinya} Markovich, T.; Andelman, D.; Podgornik, R. Charged membranes: Poisson-Boltzmann theory, 
DLVO paradigm and beyond. In \textit{Handbook of Lipid Membranes Molecular, Functional, and Materials Aspects}, ed.; CRC Press, 2021; pp 99-128. 

\bibitem{Yael2018} Avni, Y.; Markovich, T.; Podgornik, R.; Andelman, D. Charge regulating micro-ions in salt solutions: 
screening properties and electrostatic interactions. \textit{Soft Matter} $\bf{2018}$, \textit{14}, 6058-6069.

\bibitem{Flory1953} Flory, P. J. Rubber elasticity. In \textit{Principles of Polymer Chemistry}; Cornell University Press, 1953; pp 468.

\bibitem{Landsgesell2021} Landsgesell, J.; Beyer, D.; Hebbeker, P.; Ko\v sovan, P.; Holm C. The pH-dependent swelling of weak polyelectrolyte hydrogels is modeled at different levels of resolution. \textit{Macromolecules} $\bf{2022}$, \textit{55}, 3176–3188.

\bibitem{Approx} By using the identity $-a{\rm e}^{-x}+b{\rm e}^x=2\sqrt{ab}\sinh(x-\ln\sqrt{a/b})$ and rescaling the potential by subtracting the constant $\ln\sqrt{a/b}$ in all formulae, we can derive explicitly the following relation $-4\pi l_B(n_b+n_{\rm H}^b){\rm e}^{-\psi}+4\pi l_B n_b{\rm e}^{\psi}=8\pi l_B\sqrt{n_b(n_b+n_{\rm H}^b)}\sinh{\psi}=\kappa^2\sinh{\psi}$.

\bibitem{NP-regulation} Ninham, B. W.; Parsegian V. A. Electrostatic potential between surfaces bearing ionizable 
groups in ionic equilibrium with physiologic saline solution. \textit{J. Theor. Biol.} $\bf{1971}$, \textit{31},405-428.


\bibitem{Larson2016} Salehi, A.; Larson, R. G. A molecular thermodynamic model of complexation in mixtures of oppositely charged polyelectrolytes with an explicit account of charge association/dissociation. \textit{Macromolecules} $\bf{2016}$, \textit{49}, 9706-9719.

\bibitem{Zheng2021} Zheng, B.; Avni, Y.; Andelman, D.; Podgornik, R. Phase separation of polyelectrolytes: The effect of charge
regulation. \textit{J. Phys. Chem. B} $\bf{2021}$,  \textit{125}, 7863-7870.

\end{thebibliography}
\end{document}